\documentclass[twocolumn,showpacs,preprintnumbers,amsmath,amssymb,superscriptaddress]{revtex4-1}

\usepackage{graphicx}
\usepackage{dcolumn}
\usepackage{bm}

\begin{document}

\title{Comment on ``Effective thermal conductivity in thermoelectric materials'' [J. Appl. Phys. 113, 204904 (2013)]}

\author{Y. Apertet}\email{yann.apertet@gmail.com}
\affiliation{Institut d'Electronique Fondamentale, Universit\'e Paris-Sud, CNRS, UMR 8622, F-91405 Orsay, France}
\author{H. Ouerdane}
\affiliation{Laboratoire CRISMAT, UMR 6508 CNRS, ENSICAEN et Universit\'e de Caen Basse Normandie, 6 Boulevard Mar\'echal Juin, 14050 Caen, France}
\affiliation{Laboratoire Interdisciplinaire des Energies de Demain (LIED), UMR 8236 Universit\'e Paris Diderot CNRS, 5 Rue Thomas Mann, 75013 Paris France}
\author{C. Goupil}
\affiliation{Laboratoire CRISMAT, UMR 6508 CNRS, ENSICAEN et Universit\'e de Caen Basse Normandie, 6 Boulevard Mar\'echal Juin, 14050 Caen, France}
\affiliation{Laboratoire Interdisciplinaire des Energies de Demain (LIED), UMR 8236 Universit\'e Paris Diderot CNRS, 5 Rue Thomas Mann, 75013 Paris France}
\author{Ph. Lecoeur}
\affiliation{Institut d'Electronique Fondamentale, Universit\'e Paris-Sud, CNRS, UMR 8622, F-91405 Orsay, France}

\date{\today}

\begin{abstract}
In a recent article, Baranowski et al. [J. Appl. Phys. {\bf 113}, 204904 (2013)] proposed a model that allegedly facilitates optimization of thermoelectric generators operation as these latter are in contact with hot and cold temperature baths through finite conductance heat exchangers. In this Comment, we argue that the results and analyses presented by these authors are misleading since their model is incomplete and rests on an inappropriate assumption derived from thermoelectric compatibility theory.
\end{abstract}


\maketitle
\section{Introduction}
In a recent article \cite{Baranowski2013}, Baranowski and coworkers developed a model that describes heat transport inside thermoelectric generators (TEG) using an effective thermal conductivity $\kappa_{\rm eff}$. The purpose of this model is to simplify the optimization of TEGs operation, especially when dissipative thermal couplings between the generator and the heat baths are considered. In this Comment, we show that the description of this model is incomplete and suffers from some inconsistencies as conditions for optimization are applied. We thus focus on two points: first, the dependence of $\kappa_{\rm eff}$ upon the electrical working conditions; second, the use of the ``$u = s$'' condition related to the thermoelectric compatibility \cite{Snyder2003}.

\section{Influence of electrical working conditions on $\kappa_{\rm eff}$}
The introduction of an effective thermal conductivity is motivated by the need to describe the contribution of thermoelectric processes to the heat transport inside the TEG. This additional contribution is presented by Baranowski and coworkers as  \cite{Baranowski2013}:\\ 
\emph{``heat generation/consumption due to secondary thermoelectric effects''}.

\noindent We do not agree with the terminology \emph{generation/consumption} and \emph{secondary effects}. First, the term \emph{convective heat transport}, explained in the next paragraph, is more appropriate to qualify the process studied in Ref.~\cite{Baranowski2013}: the heat consumption is related to power production, which remains small compared to the heat transferred between the hot and cold reservoirs, while the heat generation is rather associated with Joule heating, which is not explicitly mentioned in the discussed article \cite{Baranowski2013}. Second, as the existence of the additional convective heat flux, compared to Fourier conduction only for non thermoelectric materials, is the main property of a TEG it should not be qualified as \emph{``secondary effect''}; quite the contrary, it is the essence of the physical phenomenon occurring in this system.

We now describe the heat transport inside the TEG using a model we developed in Ref.~\cite{Apertet2012b}. In that article, we focused on the contribution of the heat carried by the global movement of charge carriers, each one carrying a quantity of heat $e\alpha T$, with $e$ being the elementary electrical charge, $\alpha$ being the Seebeck coefficient, and $T$ being the local temperature. For sake of simplicity, this heat was assumed constant along the device and equals to $e\alpha \overline{T}$, with $\overline{T}$ being the average temperature inside the TEG. Since this process of heat transport is associated with a macroscopic motion, the electrical current $I$ was related to a \emph{convection} phenomenon \cite{Thomson1856}. Using a force-flux formalism, we expressed thus total heat flux $q_{\rm h}$ inside the TEG as the sum of a convective term and of the classical Fourier conduction:
\begin{equation}
q_{\rm h} = \underbrace{\alpha \overline{T} I}_{\rm TE~convection} + \underbrace{K {\Delta T}_{\rm TE}}_{\rm Fourier~conduction} 
\end{equation}
\noindent where ${\Delta T}_{\rm TE}$ is the temperature difference across the TEG, and $K$ is the thermal conductance of the TEG under open-circuit condition.

In order to express the effective thermal resistance $\Theta_{\rm TE}$ defined in Eq.~(8) of Ref.~\cite{Baranowski2013} as a function of the system parameters, we simply factorize the previous expression by ${\Delta T}_{\rm TE}$:
\begin{equation}\label{thetaTE}
q_{\rm h} = \left(\underbrace{\frac{\alpha \overline{T} I}{\Delta T} + K}_{1/ \Theta_{\rm TE}}\right) {\Delta T}_{\rm TE}
\end{equation}
\noindent The main interest of the above form is to highlight the dependence of $\Theta_{\rm TE}$ on the electrical working conditions. This point, while being essential to optimize the system, was overlooked in Ref.~\cite{Baranowski2013}: the derivation of $\Theta_{\rm TE}$ by Baranowski and coworkers implies that this quantity is uniquely defined and hence may lead to confusion.

As the main focus of Ref.~\cite{Baranowski2013} is the effective thermal conductivity $\kappa_{\rm eff}$, we express this quantity as a function of the load resistance $R_{\rm load}$ which reflects the dependence on the electrical working conditions. Replacing $\Theta_{\rm TE}$ in Eq.~(19) of Ref.~\cite{Baranowski2013} by the expression obtained in Eq.~(\ref{thetaTE}), we get:
\begin{equation}\label{keff}
\kappa_{\rm eff} = \kappa \left(1 + \frac{Z\overline{T}}{1+ R_{\rm load} / R_{\rm in}}\right)
\end{equation}
\noindent since the electrical current inside the TEG is $I = \alpha \Delta T/ ( R_{\rm load} +  R_{\rm in})$ with $R_{\rm in}$ being the TEG electrical internal resistance. In Eq.~(\ref{keff}), $Z = \alpha^2 / (R_{\rm in} K)$ is the TEG figure of merit. 

Thus, contrary to the statement of Baranowski and coworkers :\emph{``This model is especially powerful in that the value of $\kappa_{\rm eff}$ does not depend upon the operating conditions of the TEG but rather on the transport properties of the TE materials themselves''}, $\kappa_{\rm eff}$ varies from $\kappa$ for open circuit condition ($R_{\rm load} = \infty$) to $\kappa (1 + Z\overline{T})$ for closed-circuit condition ($R_{\rm load} = 0$). This variation however does not \emph{lessen the power} of the model. The explicit expression of the thermal properties dependence on electrical working conditions is even the key point to optimize the system: Several recent articles \cite{Apertet2012a, Yazawa2012, McCarty2012, Gomez2013, Kim2013, Castrillo2013} deal with this issue and stress importance of the interdependence of electrical and thermal optimization. For example, the optimal TEG length to maximize output power proposed in Eq.~(20) of Ref.~\cite{Baranowski2013} does not take account of electrical conditions: This expression is only valid for a specific value of the electrical load whereas Eq.~(14) derived in Ref.~\cite{Yazawa2012}, is more general as the dependence of the thermal flux on $R_{\rm load}$ is clearly explicit. 

Further, the analysis of Eq.~(\ref{keff}) allows to interpret the data plotted on Figure~3 of Ref.~\cite{Baranowski2013}: although the electrical working conditions are not explicitly given, one may assumed that, as the authors used the condition ``$u = s$'' to maximize power, $R_{\rm load} = R_{\rm in} \sqrt{1 + Z\overline{T}}$. However, when seeking power maximization, the condition corresponding to electrical resistance matching: $R_{\rm load} = R_{\rm in}$ seems more suitable, even if this relation is accurate only in the case of TEGs perfectly coupled to heat reservoirs, i.e., $\Theta_{\rm h,x} = 0$; for $\Theta_{\rm h,x} \neq 0$ this condition is slightly modified \cite{Apertet2012a} but the curve interpretation remains valid. The effective thermal conductivity is then, using Eq.~\ref{keff}, $\kappa_{\rm eff} = \kappa \left(1 + Z\overline{T}/2\right)$: the almost linear trend of the curves regarding $Z\overline{T}$ of the Figure~3 of Ref.~\cite{Baranowski2013} is then recovered. It is interesting to note that this linear dependence has already been used in Ref.~\cite{Yee2013}.

Finally, we stress that the derivations presented in this section are obtained for a constant parameters model (CPM), i.e., $\alpha$, $R_{\rm in}$ and $K$ (or, equivalently, $\kappa$) are assumed to be constant, which allows a clear description of the phenomena at stake. This model may be generalized considering local rather than global properties.

\section{Discussion on the ``$u = s$'' condition}
We showed in the previous section that the effective thermal conductivity $\kappa_{\rm eff}$ has to depend on the electrical working conditions. In Ref.~\cite{Baranowski2013}, the authors restrain their analysis to the condition ``$u = s$'', where $u$ and $s$ are respectively the relative current density and the compatibility factor as defined by the thermoelectric compatibility theory developed by Snyder and Ursell \cite{Snyder2003}. Since the relative current density is a function of the electrical current $I$, setting the value of $u$ is equivalent to defining the electrical working conditions. However, we highlight here that this choice does not correspond to the maximum power condition. The condition ``$u = s$'' is indeed associated with an efficiency maximization \cite{Snyder2003} and, despite the claim of the authors that, ``\emph{if the TE element length is allowed to vary, it can be shown that the ``$u = s$'' condition also produces the maximum power possible for the given temperature gradient and transport properties}'', power maximization is reached for an other value of the relative current density.

To illustrate this point we plot both the efficiency and the output power of a TEG as functions of the relative current density $u$ on Figure~1; the parameters are close to those used in Figure~8 of Ref.~\cite{Baranowski2013}. The TEG junction temperatures are assumed constant here but this assumption has no impact on the plot as variable $u$ is defined locally. The curve found for the efficiency is, as expected, similar that of the Figure~8 of Ref.~\cite{Baranowski2013}; its maximum corresponds to the condition ``$u = s$''. However, the maximum of the output power is obtained for $u\neq s$: It appears clearly that efficiency maximization and power maximization are two different optimization targets, which cannot be reached simultaneously, except when the TEG performances are much deteriorated, in which case both maximum efficiency and maximum power vanish \cite{Apertet2012a}. Then, by setting ``$u = s$'', Baranowski and coworkers do not maximize power contrary to what they claim.

\begin{figure}
	\centering
		\includegraphics[width=0.48\textwidth]{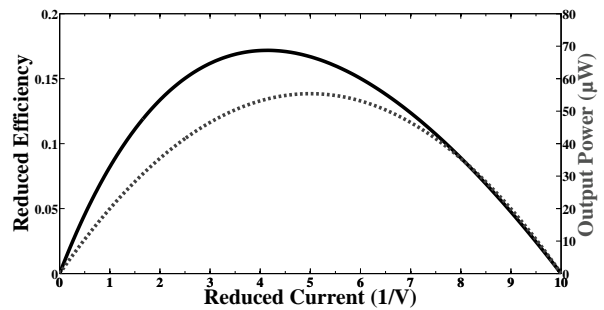}
	\caption{Reduced efficiency (straight line) and output power(dashed line) as functions of the relative current density $u$ for a TEG with $Z\overline{T} = 1$ and $\alpha \overline{T} = 0.1~V$.}
	\label{fig:figure1}
\end{figure}

\section{\bf Additional remark}
As stressed by Su and coworkers \cite{Su2013}, the ratio $\omega = \Theta_{\rm TE} / \Theta_{\rm Hx}$ should not be used as an optimization variable for the output power: this choice yields confusion since power maximization is obtained for an infinite $\omega$, if this latter variable's variations are due to $\Theta_{\rm Hx}$'s variations, with constant $\Theta_{\rm TE}$. It amounts to stating that thermal contact resistances should be minimized. The condition $\Theta_{\rm TE} = \Theta_{\rm Hx}$ corresponds to power maximization only if the thermal resistances of heat exchangers are not allowed to vary.

\section{Summary}
While the model proposed in Ref.~\cite{Baranowski2013} is valuable to the theory of TEG optimization, the analyses based on it, made by Baranowski and coworkers, are misleading as they overlook the dependence of $\kappa_{\rm eff}$ upon the electrical working conditions, in contrast with the recent literature. Moreover, the particular working condition used to maximize the output power, i.e., ``$u = s$'', actually does not correspond to power maximization but to efficiency maximization which is a distinct optimization criterion.

\end{document}